# Improved Bounds for Online Preemptive Matching

Leah Epstein[*]    Asaf Levin[†]    Danny Segev[‡]    Oren Weimann[§]

**Abstract**

When designing a preemptive online algorithm for the *maximum matching* problem, we wish to maintain a valid matching $M$ while edges of the underlying graph are presented one after the other. When presented with an edge $e$, the algorithm should decide whether to augment the matching $M$ by adding $e$ (in which case $e$ may be removed later on) or to keep $M$ in its current form without adding $e$ (in which case $e$ is lost for good). The objective is to eventually hold a matching $M$ with maximum weight.

The main contribution of this paper is to establish new lower and upper bounds on the competitive ratio achievable by preemptive online algorithms:

- We provide a lower bound of $1 + \ln 2 \approx 1.693$ on the competitive ratio of any randomized algorithm for the maximum cardinality matching problem, thus improving on the currently best known bound of $e/(e-1) \approx 1.581$ due to Karp, Vazirani, and Vazirani [STOC'90].

- We devise a randomized algorithm that achieves an expected competitive ratio of 5.356 for maximum weight matching. This finding demonstrates the power of randomization in this context, showing how to beat the tight bound of $3 + 2\sqrt{2} \approx 5.828$ for deterministic algorithms, obtained by combining the 5.828 upper bound of McGregor [APPROX'05] and the recent 5.828 lower bound of Varadaraja [ICALP'11].

## 1 Introduction

In the *maximum matching* problem, we are given an undirected graph $G = (V, E)$ whose edges have non-negative weights associated with them. A set of edges $M \subseteq E$ is called a *matching* when no two of them share a common vertex. The objective is to compute a matching of maximum total weight. Due to its wide real-life applicability, as well as to its appealing theoretical nature, this computational setting has received a great deal of attention from various communities such as computer science, mathematics, operations research, and economics (see Schrijver's book [12] and references therein for a comprehensive overview of classic work).

As can only be expected, the algorithmic research revolving around maximum matching has deviated from studying the traditional (offline) setting to other models. In particular, the online setting has been extensively studied over the last few decades [4, 5, 6, 7, 8, 9, 10]. Here, the edges (along with their weights) are presented one by one to the algorithm, which is required to keep a valid matching at all times. In other words, once an edge $e$ is presented, the algorithm must decide whether to add it to $M$ or not. However, $e$ may be added only if the resulting set of edges $M \cup \{e\}$ remains a valid matching. The online setting has two fundamental models, depending on whether the acceptance of an edge is permanent or not.

In the non-preemptive model, the decision of whether or not to add any given edge to $M$ is irrevocable, i.e., once an edge is added to the set $M$ of previously accepted edges it can never be removed. The final matching thus consists of all edges that were ever accepted. Alas, in this model, simple examples demonstrate that the competitive ratio of any (deterministic or randomized) algorithm exceeds any function of the number of vertices, meaning that no competitive algorithm exists (see for example [5]). That being said, in the unweighted case (where all edge weights are equal, which is also called the *maximum cardinality matching* problem), a greedy approach that accepts an edge whenever possible has a competitive ratio of 2. For deterministic algorithms, this ratio is actually best possible, as shown by Karp, Vazirani, and Vazirani [8].

---
[*]Department of Mathematics, University of Haifa, 31905 Haifa, Israel. Email: lea@math.haifa.ac.il.
[†]Faculty of Industrial Engineering and Management, The Technion, 32000 Haifa, Israel. Email: levinas@ie.technion.ac.il.
[‡]Department of Statistics, University of Haifa, 31905 Haifa, Israel. Email: segevd@stat.haifa.ac.il.
[§]Department of Computer Science, University of Haifa, 31905 Haifa, Israel. Email: oren@cs.haifa.ac.il.

In the preemptive model, the algorithm is given more freedom by being able to regret on (retrospectively) bad decisions, in the sense that we are now allowed to remove previously accepted edges from the current matching at any point in time; this event is called *preemption*. Nevertheless, an edge that was either rejected or preempted cannot be re-inserted to the matching later on. As opposed to the non-preemptive model, with this extra freedom competitive algorithms do exist. Specifically, a deterministic algorithm that was proposed by Feigenbaum et al. [6] attains a competitive ratio of 6. Later on, McGregor [10] improved on this finding, by tweaking it into achieving a ratio of $3 + 2\sqrt{2} \approx 5.828$. On the other hand, Epstein et al. [5] established a lower bound of 4.967 for any deterministic algorithm, which has recently been improved by Varadaraja [3] to $3 + 2\sqrt{2} \approx 5.828$.

The upper bound of McGregor and the lower bound of Varadaraja establish a tight bound of 5.828 on the competitive ratio of any *deterministic* algorithm in the preemptive model. For *randomized* algorithms, the currently best lower bound of $e/(e-1) \approx 1.581$ can be inferred from the work of Karp et al. [8] on a similar model. Their worst-case example, which will be discussed later, actually works for the unweighted case.

## 1.1 Our results

The main contribution of this paper is to establish new lower and upper bounds on the competitive ratio of *randomized* algorithms in the *preemptive model*. Our findings, along with some technical comments, can be briefly summarized as follows.

**A lower bound for unweighted graphs.** We provide a lower bound of $1 + \ln 2 \approx 1.693$ on the competitive ratio of any randomized algorithm for the maximum cardinality matching problem, thus improving on the currently best known bound of $e/(e-1) \approx 1.581$ due to Karp et al. [8] in a similar model. It is worth pointing out that the latter has originally been proven to be best possible for bipartite graphs[1], and quite surprisingly, it turns out that our construction results in a bipartite graph as well. At first glance, this seems to be a contradiction. However, in the model studied by Karp et al., the edges adjacent to each vertex of one (fixed) part of the partition are all revealed simultaneously. On the other hand, in our construction, the sequence of arriving edges no longer follows this restriction. This bound is given in Section 2.

**An upper bound for weighted graphs.** As previously mentioned, when arriving edges are accompanied by non-negative weights, there is a tight bound of $3 + 2\sqrt{2} \approx 5.828$ on the competitive ratio of any deterministic algorithm [10, 3]. An interesting open question is whether randomization offers any advantage in this context. We answer this question in the affirmative, by devising a *randomized* preemptive algorithm that beats the aforementioned bound, and achieves a competitive ratio of $\theta \approx 5.356$, where $\theta$ is the unique solution to $2(\ln \theta + 1) = \theta$ over $(2, \infty)$. This bound is given in Section 3.

---

[1]Specifically, the authors also proposed the well-known ranking algorithm, whose competitive ratio exactly matches the lower bound of $e/(e-1)$.

## 1.2 Related work

What seems to have ignited renewed interest in the preemptive online model is the investigation of maximum matching in the *semi-streaming* model, which was introduced by Muthukrishnan [11]. In this model, the algorithm is allowed to use only $O(n \cdot \text{polylog}(n))$ space at all times but is not required to hold a valid matching at all times. The possibility to keep in memory *any* set of edges (not only a matching) is what gives this model its added strength. In particular, an edge may be re-inserted to $M$, even if it has previously been removed, as long as this edge was kept in memory.

Epstein et al. [5] observed that the semi-streaming algorithms of Feigenbaum et al. [6] and McGregor [10] can actually be viewed as preemptive online algorithms in disguise. Nevertheless, the semi-streaming model is not as strict as the preemptive online model. In particular, the currently best semi-streaming algorithm for maximum weighted matching is that of Epstein et al. whose competitive ratio is 4.91. This improved on a ratio of 5.585 due to Zelke [15]. Both of these algorithms are *not* preemptive online, as the former may simultaneously hold $\Omega(\log n)$ matchings in memory (arguing that their union contains a good matching), while the latter keeps several additional edges for each edge in the current matching. When the semi-streaming algorithm is allowed to make a constant number of passes over the input stream, several algorithms of smaller approximation ratios were designed by McGregor [10] and Ahn and Guha [1, 2].

## 2  A Lower Bound for Randomized Algorithms

In this section, we establish a lower bound of $1 + \ln 2 \approx 1.693$ on the competitive ratio of any randomized algorithm in the preemptive online setting.

### 2.1  The general idea

Prior to delving into technical details, we provide a high-level description of how our construction works, along with some intuitive explanations. For ease of presentation, we use Yao's principle for profit maximization problems [13]; that is, to prove a lower bound of $C$ on the achievable (randomized) competitive ratio, we define a probability distribution on a class of inputs such that any deterministic algorithm (evaluated on this probability distribution of the inputs) must have a competitive ratio of at least $C$.

Consider a fixed deterministic online algorithm ALG. In what follows, the underlying graph will be comprised of $L$ layers (or vertical columns), each consisting of $2n$ vertices (see Figure 1). It is instructive to think of $L$ and $n$ as very large integers. With this structure in place, the input sequence starts with a random set of edges connecting vertices in layer 1 to vertices in layer 2. As soon as this sequence terminates, a new one begins, with a random set of edges between layers 2 and 3, then between layers 3 and 4, so on and so forth. The edges between adjacent layers $\ell$ and $\ell+1$ are revealed in $2n$ rounds: In each round, all the edges connecting a vertex $u$ in layer $\ell$ to its neighbors in layer $\ell+1$ are revealed one after the other. The following properties are satisfied:

**Property 1: No preemption from previous layers.** Once ALG picks a subset of edges between layers $\ell$ and $\ell+1$, there is no motivation to preempt any of them when we proceed to subsequent layers. This property is already achieved by the above description, regardless of any particular randomization method we suggest, simply due to the fact that edges are revealed layer by layer. To see this, note that if ALG preempts such edge, say $(u, v)$, then either the vertex $v$ is eventually left unmatched, meaning that we have just lost $(u, v)$ and gained nothing. Or, the vertex $v$ is matched to a vertex in layer $\ell+2$, meaning that the number of edges in the current matching remains unchanged, and we have just made things worse further down the road by (tentatively) matching a vertex in layer $\ell + 2$.

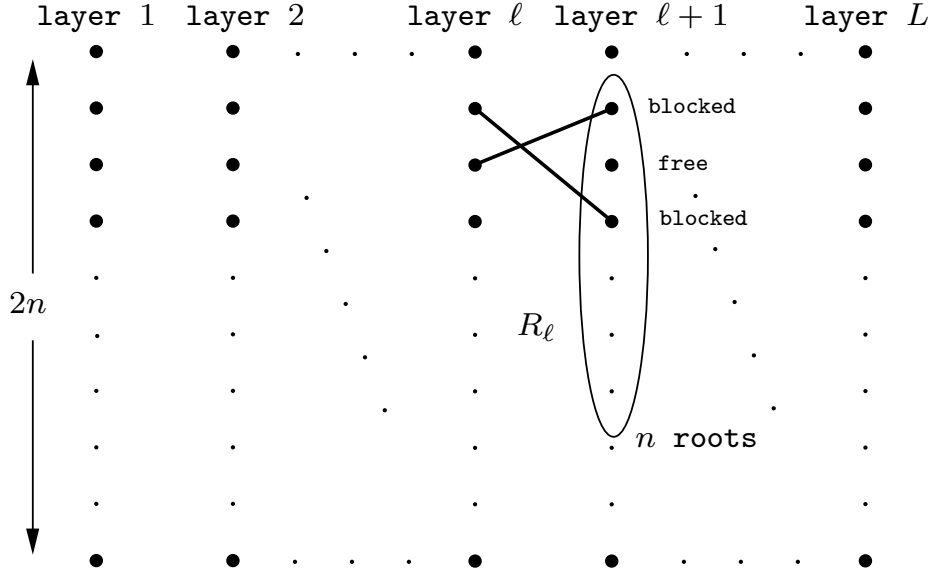

Figure 1: The layered graph used for the lower bound

**Property 2: No preemption from previous rounds.** Recall that the edges between adjacent layers $\ell$ and $\ell + 1$ are revealed in $2n$ rounds. Consider some particular round, where all the edges connecting a vertex $u$ in layer $\ell$ to its neighbors in layer $\ell + 1$ are revealed one after the other. Once this round terminates, ALG can hold a single edge $(u, v)$ for some $v$ in layer $\ell + 1$. From this point on, there is no motivation to ever preempt $(u, v)$. To verify this, note that if ALG preempts $(u, v)$, it has two options:

- Leave $v$ unmatched to other vertices in layer $\ell$, meaning that we have just lost $(u, v)$ and gained nothing, since the only possible reason to preempt this edge is to match between $v$ and vertices in layer $\ell + 2$, but there is no motivation to do so (by Property 1).

- Match $v$ to some other vertex in layer $\ell$, which leaves us at exactly the same situation when proceeding to the next layer (i.e., $v$ is still matched).

From Properties 1 and 2 it may seem that preemptions do not occur at all. However, this is certainly not the case, as preemptions can occur within a round: While the edges connecting a vertex $u$ in layer $\ell$ to all its neighbors in layer $\ell + 1$ are revealed, ALG may preempt a previously chosen edge $(u, v)$ in favor of a different one $(u, v')$.

**The bottom line.** The preceding discussion will allow us to argue that ALG picks edges in a very particular way. In fact, we are able to explicitly write the expected cardinality of the computed matching up to lower order terms, which evaluates to $\frac{1}{1+\ln 2} \cdot (L-1)n + o(Ln)$. On the other hand, in any realization of the randomized construction, we will show that there exists a feasible matching of cardinality $(L-1)n$. These observations will lead to the next theorem.

**Theorem 2.1.** *The competitive ratio of any randomized preemptive online algorithm for maximum cardinality matching is at least $1 + \ln 2$.*

## 2.2 The randomized construction

In each layer $1 \leq \ell \leq L - 1$, we will designate $n$ out of its $2n$ vertices as being *roots*, collectively forming the set $R_\ell$. These roots are a-priori unknown, and will be determined as soon as the random sequence of edges between layers $\ell - 1$ and $\ell$ terminates. This holds true for any layer other than the first, in which $R_1$ is defined by picking $n$ arbitrary vertices. Once the set of roots $R_\ell$ is determined, we linearly order $R_\ell$ by picking a random permutation of these roots, such that each of the $n!$ possible permutation is equally likely. For any $r \in R_\ell$, the (random) location of $r$ in this permutation is denoted by $L(r)$.

**Revealing the edges.** With these definitions in place, we can now explain how the edges between layers $\ell$ and $\ell + 1$ are revealed. We say that a root in $R_\ell$ is *free* when it has not been matched to some vertex in the preceding layer, $\ell - 1$; otherwise, this root is said to be *blocked*.

Based on the order determined earlier, we pick the first root $r$ and introduce edges connecting it to all $2n$ vertices in layer $\ell + 1$, which are initially colored white. If $r$ is blocked, none of these edges is picked by ALG (Property 1). Otherwise, $r$ is free and one of the newly presented edges will be picked, and will not be preempted in the future (Properties 1 and 2). We now choose a (random) vertex in layer $\ell + 1$, which is picked uniformly at random, and color it black. The second root in the linear ordering is then connected to all $2n - 1$ white vertices in layer $\ell + 1$, the same logic as to when this root is matched also applies here, and out of these $2n - 1$ white vertices one is randomly chosen to be colored black. This process continues for $n$ iterations, until all roots in $R_\ell$ have been considered[2]. We still have to define a set of roots $R_{\ell+1}$ for the next layer, and these will be the $n$ white vertices remaining in layer $\ell + 1$.

## 2.3 Analysis

Having explained how our construction works, we proceed by noting that the specific random process according to which edges are revealed can be used to derive two basic observations, comparing what can be achieved with and without knowing the input sequence in advance. These observation are summarized in the next lemma.

**Lemma 2.2.** *For $1 \leq \ell \leq L - 1$, let $F_\ell$ be a random variable that stands for the number of free roots in $R_\ell$. Then,*

1. *ALG computes a matching of expected cardinality $\sum_{\ell=1}^{L-1} \mathrm{E}[F_\ell]$.*

2. *The maximum cardinality of a matching is $(L - 1)n$.*

*Proof.* During the random process in which edges between layers $\ell$ and $\ell + 1$ are revealed, we have argued in Section 2.2 that ALG expands the matching it holds just before this process begins by adding a new edge for every free root. In other words, if $F_\ell$ out of $n$ roots in $R_\ell$ are free, ALG will add exactly $F_\ell$ edges to the matching, and these will not be preempted later on (Property 1). Therefore, evaluating the expected cardinality of the computed matching translates into evaluating the expectation of $\sum_{\ell=1}^{L-1} F_\ell$. On the other hand, the maximum cardinality of a matching in the resulting instance is $(L - 1)n$, obtained by picking, for every layer $\ell$, a matching of size $n$ between the roots $R_\ell$ and the black vertices in layer $\ell + 1$. A matching of this nature necessarily exists since the first root can be matched to the first vertex being colored black, the second root to the second vertex being colored black, and so on. □

---

[2] In particular, a free root must be connected to at least two unmatched vertices in layer $\ell + 1$. This follows by observing that in iteration $i$, the current root is connected to $2n - (i - 1)$ vertices, out of which at most $i - 1$ have previously been matched, so the number of unmatched vertices is at least $2n - 2(i - 1) \geq 2$.

**Recursively computing $\mathrm{E}[F_\ell]$.** In what follows, we derive a recursive formula that ties between the expected number of free roots in successive layers, showing that $\mathrm{E}[F_{\ell+1}]$ is close to being a linear function of $\mathrm{E}[F_\ell]$. For this purpose, we look into the question of how free roots (or equivalently, blocked roots) are created, and argue that the random permutation by which roots are processed leads to a large fraction of the next-layer roots being blocked (in expectation).

**Lemma 2.3.** *There exists a layer-independent constant $c > 0$, such that for every $1 \leq \ell \leq L-1$,*

$$|\mathrm{E}[F_{\ell+1}] - n + \ln 2 \cdot \mathrm{E}[F_\ell]| \leq c .$$

*Proof.* Let $B_{\ell+1}$ be the number of blocked roots in $R_{\ell+1}$. Since $\mathrm{E}[F_{\ell+1}] = n - \mathrm{E}[B_{\ell+1}]$, and $\mathrm{E}[B_{\ell+1}] = \mathrm{E}[\mathrm{E}[B_{\ell+1}|F_\ell]]$, it is sufficient to prove that $|\mathrm{E}[\mathrm{E}[B_{\ell+1}|F_\ell]] - \ln 2 \cdot \mathrm{E}[F_\ell]| \leq c$, for a fixed value $c > 0$, independent of $\ell$. To this end, we will show that $|\mathrm{E}[B_{\ell+1}|F_\ell = K] - \ln 2 \cdot K| \leq c$, i.e., given that there are $K$ free roots in layer $\ell$, the expected number of blocked roots in layer $\ell + 1$ is $\ln 2 \cdot K$, up to some additive constant.

Let $r_1, \ldots, r_K$ be the collection of free roots in $R_\ell$, indexed in some arbitrary order. The important observation is that blocked roots in layer $\ell+1$ are created only when they are matched to one of $r_1, \ldots, r_K$, and once this happens, survive the random recoloring step in all subsequent iterations as white vertices. For example, if some free root $r$ in layer $\ell$ is matched to $v$ in layer $\ell + 1$, and there are still $p$ remaining iterations (excluding the current one), then $v$ becomes a blocked root (at some time) with probability $\frac{n}{n+p+1}$, since out of the $n + p + 1$ current white vertices, every subset of $n$ vertices has the same probability to be the set of vertices which remain white.

Note that, for every $1 \leq k \leq K$, the location $L(r_k)$ in the random permutation is one of $1, \ldots, n$, with equal probabilities to all values. Therefore, letting $I_k$ be an indicator variable for the event where the vertex in layer $\ell + 1$ to which $r_k$ is matched survives all subsequent recoloring steps as a white vertex, we have

$$\mathrm{E}[I_k|F_\ell = K] = \Pr[I_k = 1|F_\ell = K] = \sum_{p=1}^n \frac{1}{n} \cdot \frac{n}{n+p} = H_{2n} - H_n = \ln 2 + \lambda_n ,$$

such that $|\lambda_n| = \Theta\left(\frac{1}{n}\right)$. The last equation holds since $|H_t - \ln t| = \gamma + O(\frac{1}{t})$, where $\gamma$ is the Euler-Mascheroni constant [14]. The lemma follows by observing that

$$\mathrm{E}[B_{\ell+1}|F_\ell = K] = \mathrm{E}\left[\sum_{k=1}^K I_k \middle| F_\ell = k\right] = \sum_{k=1}^K \mathrm{E}[I_k| F_\ell = k] = \ln 2 \cdot K + K\lambda_n$$

and $|K\lambda_n| = \Theta(1)$. □

**Concluding the proof of Theorem 2.1.** Recall that by Lemma 2.3, we have $\mathrm{E}[F_{\ell+1}] + \ln 2 \cdot \mathrm{E}[F_\ell] \leq n + c$ for $1 \leq \ell \leq L-1$. Taking the sum over these values of $\ell$ gives

$$\sum_{\ell=1}^{L-1} \mathrm{E}[F_{\ell+1}] + \ln 2 \cdot \sum_{\ell=1}^{L-1} \mathrm{E}[F_\ell] \leq (L-1)n + (L-1)c .$$

Using $F_L \geq 0$ and $F_1 = n$ we find

$$\sum_{\ell=1}^{L-1} \mathrm{E}[F_{\ell+1}] + \ln 2 \cdot \sum_{\ell=1}^{L-1} \mathrm{E}[F_\ell] \geq (1 + \ln 2) \cdot \sum_{\ell=1}^{L-1} \mathrm{E}[F_\ell] + F_L - F_1 \geq (1 + \ln 2) \cdot \sum_{\ell=1}^{L-1} \mathrm{E}[F_\ell] - n .$$

As a consequence, by Lemma 2.2 the expected cardinality of the matching computed by ALG is

$$\sum_{\ell=1}^{L-1} \mathrm{E}[F_\ell] \leq \frac{1}{1+\ln 2} \cdot (Ln + (L-1)c) .$$

On the other hand, the maximum cardinality of a matching in the resulting instance is $(L-1)n$, implying that the asymptotic competitive ratio of ALG (when $L$ and $n$ tend to infinity) is $1+\ln 2$. Theorem 2.1 then follows.

## 3 A Randomized Algorithm

In this section, we show that by employing randomization, the lower bound of $3 + 2\sqrt{2} \approx 5.828$ on the performance of any deterministic algorithm can be beaten. In particular, by making use of randomized geometric rounding (see, for instance, [5]), our algorithm achieves an expected competitive ratio of roughly 5.356.

### 3.1 The algorithm

**Parameters.** In what follows, we utilize two real-valued parameters: A *base* $\theta > 1$, and a *shifting value* $\phi > 0$. The base $\theta$ will be optimized later on, so that the resulting competitive ratio of our algorithm is made as small as possible. In contrast, the shifting value $\phi$ will not be fixed; instead, it will be a random variable whose value is chosen to be $\theta^\tau$, where $\tau$ is uniformly distributed over the interval $(0, 1]$.

**Weight classes and rounded weights.** We define weight classes of edges in the following way. Let $w(e)$ denote the (non-negative) weight of an edge $e$. For every $i \in \mathbb{Z}$, we let the weight class $W_i$ be the collection of edges whose weight is in the interval $[\phi\theta^i, \phi\theta^{i+1})$. Once an edge $e$ is presented, we round down its original weight $w(e)$ to the lower endpoint of the weight class it belongs to, thereby obtaining its *rounded* weight $\tilde{w}(e)$. In other words, letting $i$ be the unique integer for which $w(e) \in [\phi\theta^i, \phi\theta^{i+1})$, we set $\tilde{w}(e) = \phi\theta^i$. In the remainder of this section, the latter notation will be extended to matchings, so that $w(M)$ and $\tilde{w}(M)$ will stand for the original weight and rounded weight, respectively, of a matching $M$. In addition, for a maximum weight matching $M^*$, we denote $\text{OPT} = w(M^*)$ and $\widetilde{\text{OPT}} = \tilde{w}(M^*)$. Moreover, let $\widetilde{\text{OPT}}_\tau$ denote the profit of a maximum weight matching for the weight function $\tilde{w}$ (resulting from some choice of $\tau$).

**Maintaining a matching online.** The algorithm keeps a tentative matching $M$, which is initialized prior to reading the input sequence as $M = \emptyset$. Upon the arrival of a newly-presented edge $e = (u, v)$, we proceed as follows. Let $X(M, e)$ denote the set of edges in the matching $M$ that have a common endpoint with $e$, that is, edges that have $u$ or $v$ as an endpoint. Clearly, there could be at most two such edges. If every edge $e' \in X(M, e)$ satisfies $\tilde{w}(e') < \tilde{w}(e)$ then $e$ is inserted into $M$ while the edges in $X(M, e)$ are preempted, i.e., we set $M \leftarrow (M \setminus X(M, e)) \cup \{e\}$. Otherwise, $M$ remains unchanged.

### 3.2 Analysis

We begin by accounting for the extent to which the weight of each edge is rounded. More specifically, the next lemma shows how to evaluate the ratio between the expected rounded weight of any edge[3] to its original weight in terms of the base $\theta$. Subsequently, this allows us to bound the ratio between the OPT and $\widetilde{\text{OPT}}$.

**Lemma 3.1.** *For every edge $e$, we have $\mathrm{E}_\tau[\tilde{w}(e)/w(e)] = \frac{\theta-1}{\theta \ln \theta}$.*

*Proof.* We denote by $\tilde{w}^\tau(e)$ the value $\tilde{w}(e)$ for a given choice of $\tau$. Let $p$ be an integer and let $0 < \alpha \leq 1$ be such that $w(e) = \theta^{p+\alpha}$ is satisfied. By our definition of the weight classes $\{W_i\}_{i \in \mathbb{Z}}$,

---
[3]Note that the expectation denoted by $\mathrm{E}_\tau[\cdot]$ is taken over the random choice of the variable $\tau$.

it follows that the rounded weight $\tilde{w}^\tau(e)$ is determined by:

$$\tilde{w}^\tau(e) = \begin{cases} \theta^{p+\tau} & \text{if } \tau \leq \alpha \\ \theta^{p+\tau-1} & \text{if } \tau > \alpha \end{cases}$$

Therefore, the ratio between the expected rounded weight of $e$ to its original weight is

$$\mathrm{E}_\tau\left[\frac{\tilde{w}(e)}{w(e)}\right] = \int_0^\alpha \frac{\theta^{p+\tau}}{\theta^{p+\alpha}} d\tau + \int_\alpha^1 \frac{\theta^{p+\tau-1}}{\theta^{p+\alpha}} d\tau = \frac{1}{\ln\theta} \cdot \left(\frac{1}{\theta^\alpha}(\theta^\alpha - 1) + \frac{1}{\theta^{\alpha+1}}(\theta - \theta^\alpha)\right) = \frac{\theta-1}{\theta\ln\theta}\ .$$

□

**Lemma 3.2.** *The expected rounded weight of the optimal matching $M^*$ satisfies $\mathrm{E}_\tau[\widetilde{\mathrm{OPT}}] = \frac{\theta-1}{\theta\ln\theta}\mathrm{OPT}$. Also, for any realization of the random variable $\tau$, we have $\mathrm{OPT} < \theta \cdot \widetilde{\mathrm{OPT}}_\tau$.*

*Proof.* It is easy to verify that, due to linearity of expectation, the first claim follows from separately applying Lemma 3.1 on every edge of the optimal matching $M^*$, and summing over all edges. On the other hand, the second claim holds since, regardless of the choice of $\tau$, for any edge $e$ (in particular, those in $M^*$) we have $\tilde{w}(e)/w(e) > 1/\theta$, as the lower and upper endpoints of each weight class differ by a factor of at most $\theta$. □

Up until now we have merely discussed how the expected weight of the optimal matching $M^*$ depends on the geometric rounding procedure. We now move on to describe the technical crux in our analysis, where the online matching computed by the algorithm is compared against the rounded weight of $M^*$.

**Lemma 3.3.** *For any given value of $\tau$, the profit of the algorithm is at least $\frac{\theta-2}{2\theta-2}\widetilde{\mathrm{OPT}}$.*

*Proof.* We consider how the algorithm operates for an arbitrary choice of the variable $\tau$. For this purpose, consider an optimal solution $\tilde{M}$ for the "rounded" instance, i.e., a maximum weight matching with respect to the rounded weights $\tilde{w}(\cdot)$. Clearly, $\tilde{w}(\tilde{M}) \geq \tilde{w}(M^*) = \widetilde{\mathrm{OPT}}$, as $M^*$ is an optimal matching for the original weights, but not necessarily for the rounded weights.

We associate every edge of $\tilde{M}$ with an edge in the matching $M$ that is being held by the algorithm upon termination, possibly using a single edge in $M$ as a target for several edges in $\tilde{M}$. The way this association will be defined later on enables us to argue that the ratio between the total rounded weight of edges associated with any edge $e \in M$ and between $\tilde{w}(e)$ is at most $\frac{2\theta-2}{\theta-2}$. This claim immediately leads to a ratio of $\frac{2\theta-2}{\theta-2}$ between $\widetilde{\mathrm{OPT}}$ and the total profit of the algorithm according to the rounded weights $\tilde{w}(\cdot)$. Since $w(e) \geq \tilde{w}(e)$ for every edge $e$, the actual profit of the algorithm can only be higher.

For every edge $e \in M$, we create a *preemption tree*. The root of this tree is $e$, and the children of an edge $e'$ are all edges that were preempted from the matching kept by the algorithm when $e'$ was inserted. Note that the number of children is either 2, 1, or 0, since any matching cannot contain more than a single edge for every endpoint of $e'$ (prior to the arrival of $e'$). The set of edges that do not belong to any preemption tree are edges that the algorithm never accepted, while the union of all edge sets over all trees is exactly the set of edges that belonged to the matching at some point in time.

By definition of the algorithm, for every edge $\tilde{e}$ that does not belong to a tree, there exists an edge $e'$ that does belong to a tree, sharing an endpoint with $\tilde{e}$, such that $\tilde{w}(e') \geq \tilde{w}(\tilde{e})$. If for an edge $\tilde{e} \in \tilde{M}$ such that $\tilde{e}$ does not belong to a tree, there exists a unique such edge $e'$, then $\tilde{e}$ is associated with the root of the tree where $e'$ appears. Otherwise, one such edge $e'$ is chosen arbitrarily, and $\tilde{e}$ is associated with the root of the tree of $e'$ in this case as well. For an edge $\tilde{e} \in \tilde{M}$ that belongs to some tree, $\tilde{w}$ is associated with the root of the tree in which it appears.

Next, consider a specific tree with the root $e \in M$. For an edge $\hat{e}$ of distance $d[\hat{e}]$ from the root, measured in number of edges, let $\tilde{w}_d(\hat{e}) = \tilde{w}(\hat{e})/\theta^{d[\hat{e}]}$.

**Claim 3.4.** *For every edge $\hat{e}$ in the tree, $\tilde{w}(\hat{e}) \leq \tilde{w}_d(\hat{e})$.*

*Proof.* Consider the path $e_0, e_1, e_2, \ldots, e_{d[\hat{e}]}$, where $e_0 = \hat{e}$, $e_{d[\hat{e}]} = e$, and for every $1 \leq i \leq d[\hat{e}]$, $e_i$ is the edge whose arrival caused $e_{i-1}$ to be preempted. By definition of the algorithm, $\tilde{w}(e_i) > \tilde{w}(e_{i-1})$. Since our geometric rounding method guarantees that, when rounded weights are not identical, they differ by a multiplicative factor of at least $\theta$, it follows that $\tilde{w}(e_i) \geq \theta \tilde{w}(e_{i-1})$. Therefore, $\tilde{w}(e) = \tilde{w}(e_{d[\hat{e}]}) \geq \theta^{d[\hat{e}]} \tilde{w}(e_0) = \theta^{d[\hat{e}]} \tilde{w}(\hat{e})$, or alternatively $\tilde{w}(\hat{e}) \leq \tilde{w}_d(\hat{e})$. □

For a vertex $v$ in the graph, and for every possible distance $d \geq 0$, we say that $v$ is *new* for $d$ if: (1) there is an edge of depth $d$ in some preemption tree for which $v$ is an endpoint; and (2) $v$ is not an endpoint of any edge of smaller depth. Also, The two endpoints of an edge $e \in M$ are new for $d = 0$ and are not new for any other value of $d$.

**Claim 3.5.** *For every $d > 0$, there are at most $2^d$ new vertices d. Moreover, every edge in a tree in level $d > 0$ has at most one new vertex.*

*Proof.* Since every edge has at most two children, the number of edges of depth $d$ is at most $2^d$. Every such edge has a common endpoint with an edge of depth $d - 1$, thus there is at most one new vertex per edge, which results in a total of at most $2^d$ new vertices. □

**Claim 3.6.** *For every edge $e \in M$, the total rounded weight of the edges associated with $e$ is at most $\frac{2\theta-2}{\theta-2}\tilde{w}(e)$.*

*Proof.* For an edge $\tilde{e} \in \tilde{M}$, let $d_{\min}[\tilde{e}]$ be the minimum depth such that an edge of this depth has a common vertex with $\tilde{e}$. By definition, this common vertex must be new for level $d_{\min}[\tilde{e}]$. Let $\bar{e}$ be the edge of level $d_{\min}[\tilde{e}]$ with the common vertex. Since there exists an edge in the tree having a common endpoint with $\tilde{e}$ of rounded weight at least $\tilde{w}(\tilde{e})$, we have $\tilde{w}(\bar{e}) \geq \tilde{w}(\tilde{e})$. In addition, since $\tilde{M}$ is a matching, for every edge in the tree and every new vertex $v$ that edge, at most one edge of $\tilde{M}$ has $v$ as an endpoint. By Claim 3.5, every edge of the tree other than $e$ has at most one new vertex, so the total rounded weight is at most the total rounded weight of all edges in the tree plus $\tilde{w}(e)$. Letting $\mathcal{T}(e)$ denote the set of edges in the tree rooted at $e$, by Claim 3.4 it follows that the total rounded weight is at most

$$\sum_{\bar{e}\in\mathcal{T}(e)} \tilde{w}(\bar{e}) + \tilde{w}(e) \leq \sum_{\bar{e}\in\mathcal{T}(e)} \tilde{w}_d(\bar{e}) + \tilde{w}(e) \leq \sum_{d=0}^{\infty} 2^d \frac{\tilde{w}(e)}{\theta^d} + \tilde{w}(e) \leq \left(\frac{1}{1-2/\theta} + 1\right)\tilde{w}(e) \ .$$

□

This completes the proof of Lemma 3.3. □

**Theorem 3.7.** *The algorithm achieves an expected competitive ratio of $\frac{2\theta \ln \theta}{\theta-2}$. This ratio is minimized for $\theta^* \approx 5.356$, where $\theta^*$ is the unique solution to $2(\ln \theta + 1) = \theta$ over $(2, \infty)$, in which case its value is $\theta^*$.*

*Proof.* By Lemmas 3.2 and 3.3, we have $\mathrm{E}_\tau[w(M)] \geq \frac{\theta-2}{2\theta-2}\mathrm{E}_\tau[\widetilde{\mathrm{OPT}}] \geq \frac{\theta-2}{2\theta-2} \cdot \frac{\theta-1}{\theta \ln \theta}\mathrm{OPT} = \frac{\theta-2}{2\theta \ln \theta}\mathrm{OPT}$. □

**References**

[1] K. J. Ahn and S. Guha. Laminar families and metric embeddings: Non-bipartite maximum matching problem in the semi-streaming model. Available online at: http://arxiv.org/abs/1104.4058.

[2] K. J. Ahn and S. Guha. Linear programming in the semi-streaming model with application to the maximum matching problem. In *Proceedings of the 38th International Colloquium on Automata, Languages and Programming (ICALP)*, pages 526–538, 2011.

[3] A. Badanidiyuru Varadaraja. Buyback problem – approximate matroid intersection with cancellation costs. In *Proceedings of the 38th International Colloquium on Automata, Languages and Programming (ICALP)*, pages 379–390, 2011.

[4] N. Bansal, N. Buchbinder, A. Gupta, and J. Naor. An $O(\log^2 k)$-competitive algorithm for metric bipartite matching. In *Proceedings of the 15th annual European Symposium on Algorithms (ESA)*, pages 522–533, 2007.

[5] L. Epstein, A. Levin, J. Mestre, and D. Segev. Improved approximation guarantees for weighted matching in the semi-streaming model. *SIAM Journal on Discrete Mathematics*, 25(3):1251–1265, 2011.

[6] J. Feigenbaum, S. Kannan, A. McGregor, S. Suri, and J. Zhang. On graph problems in a semi-streaming model. *Theoretical Computer Science*, 348(2-3):207–216, 2005.

[7] B. Kalyanasundaram and K. Pruhs. Online weighted matching. *Journal of Algorithms*, 14(3):478–488, 1993.

[8] R. Karp, U. Vazirani, and V. Vazirani. An optimal algorithm for on-line bipartite matching. In *Proceedings of the 22nd Annual ACM Symposium on Theory of Computing (STOC)*, pages 352–358, 1990.

[9] S. Khuller, S. Mitchell, and V. Vazirani. On-line algorithms for weighted bipartite matching and stable marriages. *Theoretical Computer Science*, 127(2):255–267, 1994.

[10] A. McGregor. Finding graph matchings in data streams. In *Proceedings of the 8th International Workshop on Approximation Algorithms for Combinatorial Optimization Problems (APPROX)*, pages 170–181, 2005.

[11] S. Muthukrishnan. *Data Streams: Algorithms and Applications*. Foundations and Trends in Theoretical Computer Science. Now Publishers Inc, 2005.

[12] A. Schrijver. *Combinatorial Optimization: Polyhedra and Efficiency*, volume 24 of *Algorithms and Combinatorics*. Springer, 2003.

[13] A. C.-C. Yao. Probabilistic computations: Toward a unified measure of complexity (extended abstract). In *Proceedings of the 18th Annual Symposium on Foundations of Computer Science (FOCS)*, pages 222–227, 1977.

[14] R. M. Young. Euler's constant. *The Mathematical gazette*, 75:187–190, 1991.

[15] M. Zelke. Weighted matching in the semi-streaming model. *Algorithmica*, 62(1-2):1–20, 2012.